\documentclass[apjl,iop]{emulateapj}
\usepackage[utf8]{inputenc}
\usepackage{natbib} 
\usepackage{color}

\newcommand{\referee}[1]{#1}

\begin{document}

\title{First images of debris disks around TWA~7, TWA~25, HD~35650, and HD~377}

\author{
	\'Elodie Choquet\altaffilmark{1}, 
	Marshall D. Perrin\altaffilmark{1}, 
	Christine H. Chen\altaffilmark{1}, 
	R\'emi Soummer\altaffilmark{1}, 
	Laurent Pueyo\altaffilmark{1}, 
	James B. Hagan\altaffilmark{1,2},    
	Elena Gofas-Salas\altaffilmark{1,3},  
	Abhijith Rajan\altaffilmark{4},   
	David A. Golimowski\altaffilmark{1},            
	Dean C. Hines\altaffilmark{1}, 
	Glenn Schneider\altaffilmark{5},             
	Johan Mazoyer\altaffilmark{1},    
	Jean-Charles Augereau\altaffilmark{6,7},   
	John Debes\altaffilmark{1},         
	Christopher C. Stark\altaffilmark{1},           
	Schuyler Wolff\altaffilmark{8},        
	Mamadou N'Diaye\altaffilmark{1},     
	\referee{Kevin Hsiao}\altaffilmark{8}
}
\altaffiltext{1}{Space Telescope Science Institute, 3700 San Martin Dr, Baltimore MD 21218, USA}
\altaffiltext{2}{Purdue University, West Lafayette IN 47907, USA}
\altaffiltext{3}{Institut d'Optique Graduate School, 2 Avenue Augustin Fresnel, 91127 Palaiseau, France}
\altaffiltext{4}{Arizona State University, Phoenix AZ 85004, USA}
\altaffiltext{5}{Steward Observatory, The University of Arizona, 933 North Cherry Avenue, Tucson, AZ 85721, USA}
\altaffiltext{6}{Univ. Grenoble Alpes, IPAG, F-38000 Grenoble, France}
\altaffiltext{7}{CNRS, IPAG, F-38000 Grenoble, France}
\altaffiltext{8}{Johns Hopkins University, 3400 North Charles Street, Baltimore MD 21218, USA}

\email{Contact: choquet@stsci.edu}

\begin{abstract}
We present the first images of four debris disks observed in scattered light around the young (4--250~Myr old) M dwarfs TWA~7 and TWA~25, the K6 star HD~35650, and the G2 star HD~377. We obtained these images by reprocessing archival Hubble Space Telescope NICMOS coronagraph data with modern post-processing techniques as part of the Archival Legacy Investigation of Circumstellar Environments (ALICE) program. All four disks appear faint and compact compared with other debris disks resolved in scattered light. The disks around TWA~25, HD~35650, and HD~377 appear very inclined, while TWA~7's disk is viewed nearly face-on.  The surface brightness of HD~35650's disk is strongly asymmetric. These new detections raise the number of disks resolved in scattered light around M and late-K stars from one (the AU Mic system) to four. This new sample of resolved disks enables comparative studies of heretofore scarce debris disks around low-mass stars relative to solar-type stars.
\end{abstract}


\section{Introduction}

Debris disks are second-generation dust belts orbiting main sequence stars after their periods of planet formation, and are counterparts to our own solar system's asteroid and Kuiper belts. They are found around stars as young as $\sim10$~Myr old and as as old as several Gyr \citep{Hillenbrand2008}. Although small dust particles are rapidly cleared from stellar environments by Poynting-Robertson drag, radiative pressure, and stellar wind, debris disks are relatively long-lived because they are constantly replenished by destructive collisional cascades from planetesimals to micron-size particles \citep{Wyatt2008,Krivov2010}. 

Infrared surveys have shown that 24\% of A-stars and 20\% of F, G, and K stars show excess emission indicating the presence of a cold debris belt \citep{Thureau2014,Eiroa2013}.  On the other hand, mid-infrared and millimeter surveys of low-mass stars have debris-disk detection rates of only 5\% \citep{Plavchan2009,Lestrade2009}. This trend toward fewer debris disks around late-type stars is also observed by infrared interferometric searches for warm asteroid belts \citep{Absil2013,Ertel2014,Mennesson2014}.  These statistics may be biased by current sensitivity limits, however, as disks around cool late-type stars are themselves cooler and thus fainter and emitting at longer wavelengths than their solar-type counterparts.

Debris disks also provide evidence of earlier stages of planet formation. Planets may stir the planetesimal belts and thus enhance the production of small dust grains. A planet may also imprint itself in a disk's morphology even if the planet itself lies below instrumental detection limits.
Comparative debris disk studies provide opportunities to better understand the formation, composition, and evolution of planetary systems, providing a large number of disk are spatially resolved.
Unfortunately, their faintness and proximity to bright host stars make debris disks hard to image. 

\begin{figure}
\center
\epsscale{1.15}
\plotone{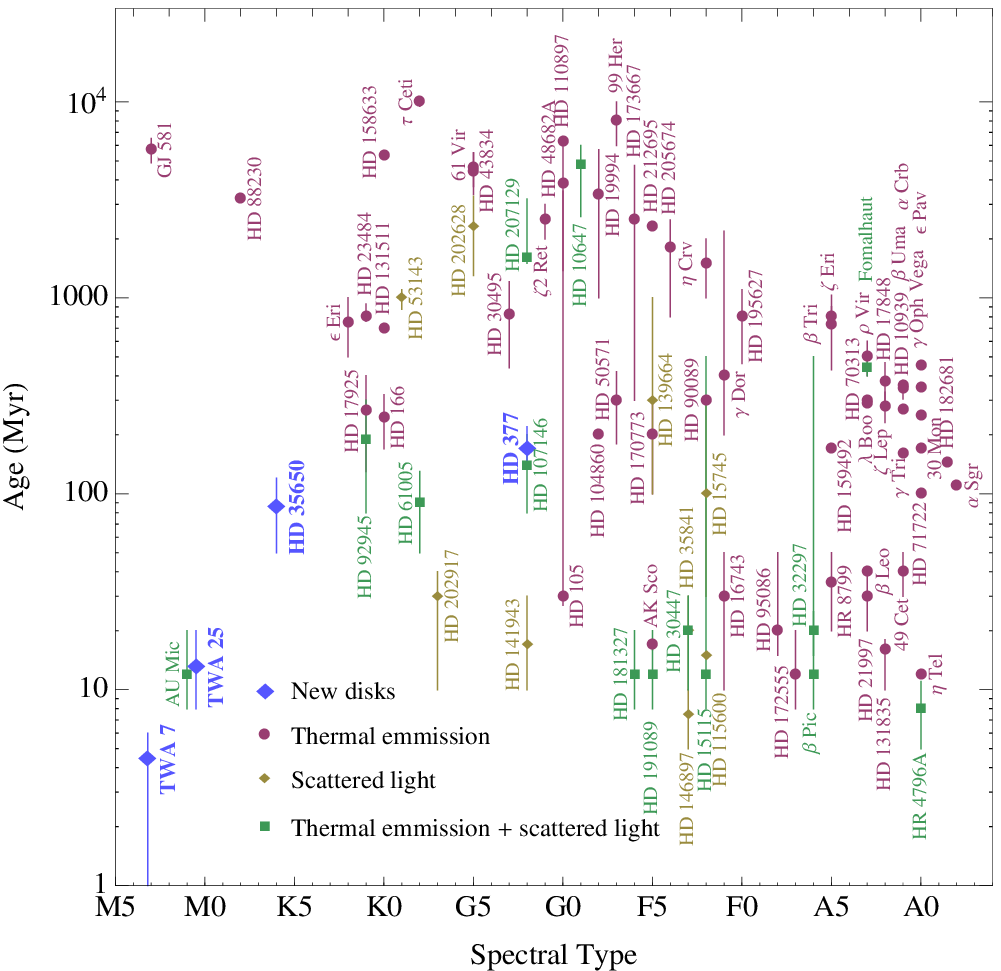}
\figcaption{
Age and spectral type of the 84 stars with resolved debris disks. Purple circles, gold diamonds, and green squares show stars with a disk resolved respectively in thermal emission only, in scattered light only, and both in thermal emission and scattered light. Large blue symbols with boldface names show the stars from this paper.
\label{Fig:AgeVsSpecType}}
\end{figure}

About 80 debris disks have been resolved around nearby stars, the majority of which surround A, F, and G stars (Fig.~\ref{Fig:AgeVsSpecType}).  Only three disks have been previously imaged around M and late-K stars: AU Mic \citep{Kalas2004}, GJ~581 \citep{Lestrade2012}, and HD~88230 \citep{Eiroa2013}, and only AU Mic's disk has been imaged in scattered light.  Most resolved debris disks have been seen only in thermal emission \citep[cf. the DUNES and DEBRIS surveys][]{Eiroa2013,Booth2013}, and only 26 in  scattered light, including the four presented in this paper.  Resolved images are necessary to constrain the morphology and dust composition of debris disks, which are degenerate parameters in spectral energy distribution (SED) model fitting. Scattered-light images also constrain the scattering properties (albedo, phase function) of the dust, which further constrain its composition.

In this letter, we present Hubble Space Telescope (HST) NICMOS scattered-light images of four debris disks that were previously unresolved. Three disks surround the low-mass stars TWA~7, TWA~25, and HD~35650, and the fourth is seen around a young solar analog, HD~377. \referee{These new debris disk detections are the result of our consistent reanalysis of the 408 targets in the NICMOS coronagraphic archive within a $6\arcsec\times6\arcsec$ field-of-view using advanced PSF-subtraction techniques. They complement the five debris disks previously reported by our group in \citet{Soummer2014}.}

\section{Dataset and data reduction}\label{sec:data}

\subsection{Archival Datasets}

The images presented in this letter come from reprocessed archival NICMOS datasets originally obtained by HST programs 7226 (PI E. Becklin), 10176 (PI I. Song), and 10527 (PI. D. Hines) (Table~\ref{tab:HST}). The first two programs were surveys for planets around young nearby stars, and the last one looked for debris disks around main-sequence solar-type stars with infrared excesses detected by the Spitzer Space Telescope.

NICMOS is an HST infrared instrument that operated between 1997 and 2008. Its NIC2 channel features a coronagraph with a 0\farcs3-radius occulting spot, a classical Lyot stop, and a spatial sampling of 0\farcs0757 pixel$^{-1}$.  The reprocessed archival images were obtained with the F110W filter (about twice as wide as the J band) or F160W filter (similar to the H band). All four targets were observed at two orientations of the telescope separated by $\sim30\degr$, as was routinely done during HST coronagraphic observations in order to remove the residual star point spread function (PSF) in the images via roll subtraction.

\begin{deluxetable}{lcccc}
\tablecaption{HST observations of the targets.\label{tab:HST}}
\tablehead{
\colhead{Target} & \colhead{Program} & \colhead{Filter} &\colhead{Date}  &\colhead{Tot. exp. time}\\
\colhead{}	&	\colhead{}		&\colhead{}	&\colhead{(UT)}		&\colhead{(s)}}
\startdata
TWA~7 		& 7226	& F160W\tablenotemark{a}	& 1998-03-26 	&1215.76\\
TWA~25		& 10176 	& F160W\tablenotemark{a} 	& 2005-02-07 	&1503.57 \\
HD~35650 	& 10176 	& F160W\tablenotemark{a} 	& 2004-10-28  	&1503.57\\
HD~377 		& 10527  	& F110W\tablenotemark{b} 	& 2005-09-16 	&4319.38
\tablenotetext{a}{Pivot wavelength $\lambda_p=1.123~\mu$m, bandpass full width at half maximum $FWHM=0.527~\mu$m}
\tablenotetext{b}{Pivot wavelength $\lambda_p=1.603~\mu$m, bandpass full width at half maximum $FWHM=0.401~\mu$m}
\enddata
\end{deluxetable}

\subsection{Data Reduction\label{Sec:datared}}

The data were reprocessed as part of the Archival Legacy Investigation of Circumstellar Environments (ALICE) project, a comprehensive reanalysis of the NICMOS coronagraphic archive using advanced post-processing methods. The images were previously recalibrated  as part of the Legacy Archive PSF Library And Circumstellar Environments (LAPLACE) program \citep{Schneider2010}.  

To subtract the residual starlight, the data were processed with the KLIP algorithm \citep{Soummer2012} based on principal component analysis (PCA) of a library of PSFs. We used the ALICE reduction pipeline \citep{Choquet2014d} to (1) assemble the libraries, (2) register the images  by minimizing their total squared difference with a reference image within a masked region containing only the diffraction pattern from the telescope spiders, (3) subtract synthetic PSFs generated with KLIP from the target images, and (4) combine all images weighted by their exposure times after rotation to North up. The final images are shown in column 1 of Fig.~\ref{Fig:4disks}.

Once a disk is identified, several reduction parameters are visually optimized to find the best compromise between speckle noise suppression and throughput maximization: the composition and size of the libraries, and the number of subtracted modes. We assembled a PSF library with reference star images only for TWA~7 which reveals a nearly face-on disk, using the Reference star Differential Imaging \referee{\citep[RDI,][]{Lafreniere2009}} strategy (258 images, 41 stars). For the three other targets, we assembled  PSF libraries with reference star images and included the images of the target in the complementary telescope orientation, thus combining Angular Differential Imaging \referee{\citep[ADI,][]{Marois2006}} with the RDI strategy (601 images from 75 stars for TWA~25 and HD~35650, and 820 images from 73  stars for HD~377). 
The ADI-type images make up less than 1\% of the libraries and help removing the residual starlight from the images while minimizing self-subtraction artifacts.
The synthetic PSFs were built out of the 125, 365, 363, and 631 highest principal components of the libraries respectively for  TWA~7, TWA~25, HD~35650, and HD~377. 
These parameters correspond to an aggressive reduction process which is necessary to distinguish these very faint disks from the starlight. \referee{It creates negative features in the images, which are faithfully reproduced and calibrated in the forward modeling process described in Sec.~\ref{sec:FWM}.}

Signal-to-noise ratio (S/N) per resolution element maps are presented in column 2 of Fig.~\ref{Fig:4disks}. The noise maps were calculated with the same method as in \citet{Soummer2014}, by computing the pixel-to-pixel noise covariance matrix of a statistically representative set of reference images reprocessed consistently with the science targets, then estimating the noise variance at each pixel within the same aperture as used for signal extraction (one resolution element in diameter).

The almost edge-on disks around TWA~25, HD~35650, and HD~377 are identified in each individual roll image, which makes the detection unambiguous. The disk around TWA~7 is also identified in the two roll images, but the rotation of the disk between the two rolls is hardly detectable because of the low inclination of the disk.

\section{Results}\label{sec:results}

\subsection{Disk detections}\label{sec:detections}

\begin{deluxetable*}{l|ccc|cc|ccc}
\tablecaption{System properties\label{tab:system}}
\tablehead{
\colhead{Name}&\colhead{SpT} 	& \colhead{$V$} 	& \colhead{$L_{IR}/L_\star$} 	& \colhead{Dist.} &\colhead{Ref.}  	&\colhead{Association} 	& \colhead{Age} 	&  \colhead{Ref.}\\
			&  				& 				& \colhead{($\times10^{-4}$)} & \colhead{(pc)} & 				&  					& \colhead{(Myr)} 	&
}
\startdata
TWA~7 		& M3.2 	& 10.91 	& 22 		& $34.5 \pm 2.5$ 	& 1		& TWA 	    & 4.4--13 	    & 2, 3	 	\\
TWA~25		& M0.5 	& 11.41 	&  \nodata	& $54 \pm 4$ 		& 4 	& TWA 	    & 7--13		& 	2, 3 	\\
HD~35650 	& K6V 	& 9.054 	& 1.75 		& $18.0 \pm 0.3$	& 5		& AB Dor 	&50--200        & 3, 6	\\
HD~377 		& G2V	& 7.59 	    & 2.0 		& $39.1 \pm 1.1$ 	& 5		& Field	    & 147--250 	    & 7, 8, 9
\tablerefs{(1) \citet{Ducourant2014}; (2) \citet{Herczeg2014}; (3) \citet{Bell2015}; (4) \citet{Weinberger2013}; (5) \citet{vanLeeuwen2007}; (6) \citet{Malo2013}; (7) \citet{Wright2004}; (8) \citet{Isaacson2010}; (9) \citet{Vican2014}}
\enddata
\end{deluxetable*}

\subsubsection{TWA~7}

TWA 7 is a young M3.2 star at a distance of $34.5\pm2.5$~pc. It is associated with the TW Hydrae moving group which has an isochronal age estimated to $10\pm3$~Myr \citep{Bell2015}, although another recent study estimated a discrepant  age of 4~Myr for TWA~7 \citep{Herczeg2014}.
No emission line is detected in the Herschel-PACS spectrum of the system, and the H$_\alpha$ line width agrees with no on-going accretion, which both indicate that the disk is an evolved second-generation debris disk \citep{Riviere-Marichalar2013}. 
It has a significant infrared excess detected from 24~$\mu$m to sub-millimeter wavelengths \citep{Low2005,Matthews2007}, best fitted by  two dust populations with a warmer inner disk at 66~K at 38~AU from the star, and a cold outer disk at 20~K and 75~AU \citep{Riviere-Marichalar2013}. 

Our NICMOS image reveals an inclined disk  35~AU ($\sim1$\arcsec) from the star.  The disk is detected with an S/N between 2 and 8 per resolution element, and S/N$\sim43$ integrated over the disk surface. \referee{Although the observed radius is compatible with the inner disk inferred from the SED analysis by \cite{Riviere-Marichalar2013}, we cannot assert that the observed disk is the innermost one without  fitting models to both the SED and the resolved image of the system, since dust properties and radius are degenerate parameters in SED-only fitting}. We found no evidence of a secondary outer dust belt, which might be too faint or oversubtracted. 
The point source at 2.4\arcsec{} from the star is a background star reported in \citet{Lowrance2005}.

\subsubsection{TWA~25}

TWA 25 is a M0.5 star at a distance of 54~pc. It is a member of the TW Hydrae association, and has an age estimated to 13~Myr \citep{Herczeg2014}, consistent with the moving group age estimation from \citet{Bell2015}. TWA~25 age and spectral type make it a near twin of AU Mic. 
No infrared excess was detected in Herschel-PACS photometry\citep{Riviere-Marichalar2013}. It has however an  excess in WISE 3 but none in  WISE 1 nor WISE 4 \citep{Schneider2012}. 
Yet our NICMOS image shows a very inclined disk at a radius of 78 AU ($\sim1.5$\arcsec). The disk is detected with an S/N between 2 to 7 per resolution element, and S/N$\sim18$ integrated over its surface.

\subsubsection{HD~35650}

HD~35650 is a K6V star at 18 pc associated with the AB Doradus moving group \citep{Malo2013} which has an age estimated between 50 and 200~Myr \citep{Malo2013,Bell2015}. 
Spitzer observations showed that the system has no infrared excess at 24~$\mu$m but detected significant excess at 70~$\mu$m indicative of a debris disk, and fitted with  dust  at temperature 60~K \citep{Zuckerman2011}. Our NICMOS image shows an asymmetric edge-on disk of radius 54~AU (2.8\arcsec{}), with the northern side of the disk $\sim 70\%$ brighter than the southern side. The star proper motion might partially explain this strong asymmetry. The disk might be compressed along its North-West axis by moving into a dense interstellar medium with a relative speed of $\sim6$~km/s, which is assumed to be a common reason for debris disks asymmetries \citep{Debes2009}. The disk is detected at a moderate S/N  $\sim2-4$ per resolution element, and S/N$\sim19$ integrated over the disk surface.

\subsubsection{HD~377}

HD 377 is G2V field star with an age estimated between 147 and 250~Myr by gyrochronology and chromospheric activity analysis \citep{Wright2004,Isaacson2010,Vican2014}, at a distance of 39.1~pc. It has infrared excesses detected from 24~$\mu$m to 160~$\mu$m with Spitzer \citep{Hillenbrand2008}, at 22~$\mu$m with WISE \citep{Vican2014}, and at 850~$\mu$m with JCMT SCUBA-2 \citep{Panic2013}. 
These excesses are best fitted by two dust populations \citep{Chen2014}, with a warm belt at 240~K at 2.4~AU and a cold dust belt at 57~K at 136.4~AU from the star. 
Our NICMOS image shows an edge-on disk at a radius of 86~AU ($\sim2.2$\arcsec). The disk is detected with S/N between 2 and 8 per resolution element, with an S/N of 12 integrated over the disk surface.

\begin{figure*}
\center
\plotone{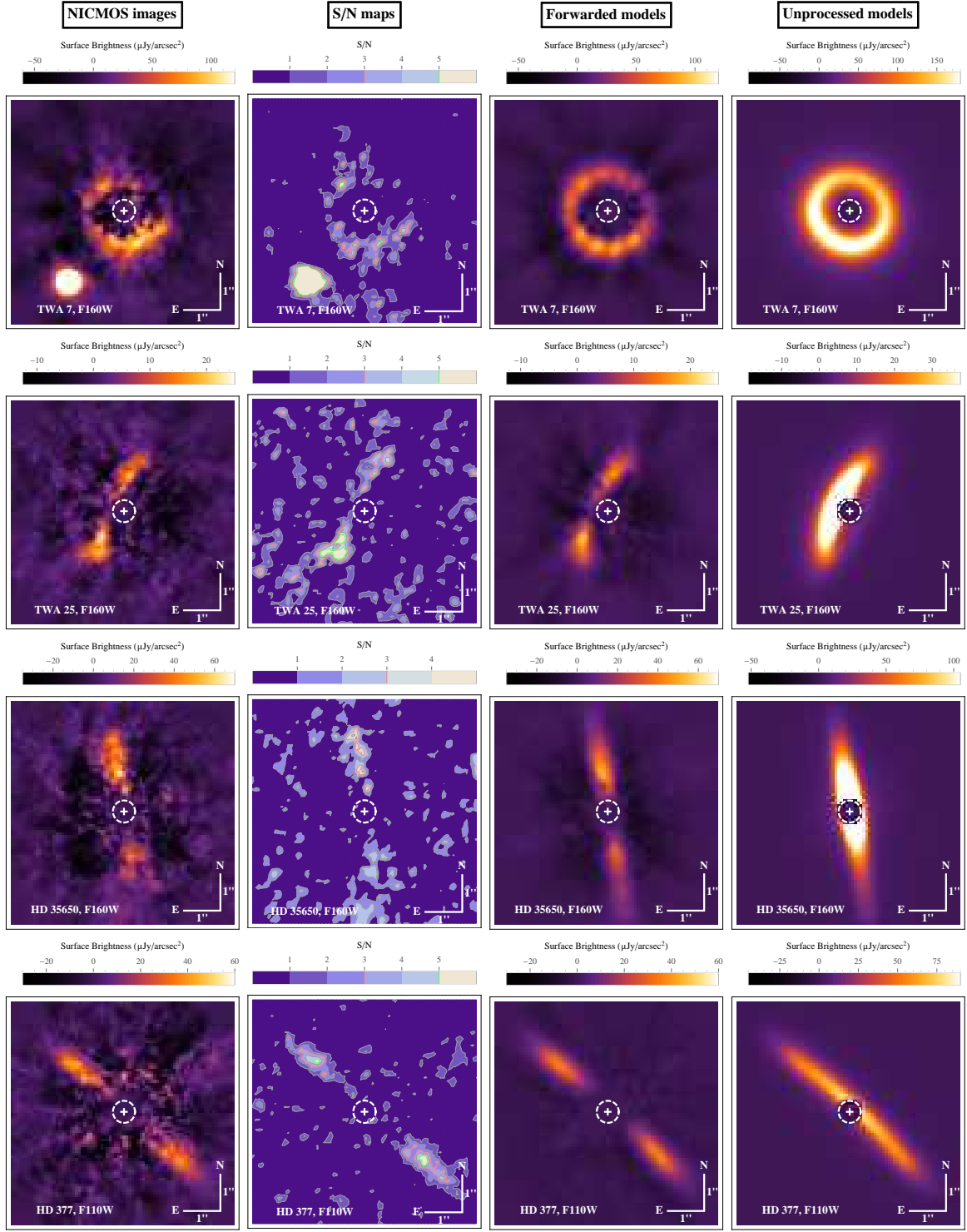}
\figcaption{From left to right: scattered-light images, S/N maps, best-fit forwarded models, and input unprocessed models of the four debris disks newly resolved from archival HST-NICMOS coronagraphic data. The NICMOS images and the forwarded models have been smoothed by convolution with a synthetic PSF. The white crosses and dashed circles show the position and size of the coronagraphic mask. The red and green lines in the S/N maps respectively trace the 3$\sigma$ and 5$\sigma$ contours.\label{Fig:4disks}}
\end{figure*}

\subsection{Photometry of the disks}\label{sec:FWM}

PSF subtraction with modern algorithms is an aggressive process that can significantly bias the morphology and photometry of disks, especially at short separations from the star, where an edge-on disk may get subtracted along with the stellar PSF component at that location. 
These effects can be  calibrated for targets processed with a PCA-based algorithm using forward modeling \citep{Soummer2012}. To calibrate the photometry of each debris disk, we (1) created models of the disk convolved with a synthetic unocculted NICMOS PSF generated with the Tiny TIM software package \citep{Krist2011} and with as many images and the same field of view orientations as each target dataset, (2) propagated the model images onto the same principal components of the PSF library as were used for the target images, (3) subtracted the part over-subtracted from the model images, and (4) derotated and combined the images as done for the NICMOS data. This ensures that biases related to interpolations necessary for image rotations are also calibrated.  The final forward-modeled images are scaled to have the same flux as the observed disks. The disk fluxes are measured by aperture photometry and are corrected by the local background levels estimated in areas close to -- but excluding -- the disks. 

Basic models reproducing the observed brightness distribution of the disks are sufficient to estimate their total flux and surface brightness. 
Detailed disk analyses including dust grain properties are out of the scope of this paper and will be studied in subsequent publications. We used the GRaTer radiative transfer code \citep{Augereau1999a,Lebreton2012} to create scattered light images of optically thin debris disks. We used a Gaussian profile for the vertical distribution of the dust density, with a linear vertical scale height of opening angle $h=10$\%. The radial distribution is simulated by a combination of two power laws inward and outward from the parent population belt at radius $r_0$, set to moderately steep slopes $\alpha_{in}=5$ and $\alpha_{out}=-5$ respectively, \referee{thus simulating a ring profile. We additionally simulated a filled disk with no gap ($\alpha_{in}=0$) for TWA~7 whose geometry might enable to constrain its inner radial profile}. The images of the disk models are created assuming an anisotropic \citet{Henyey1941} scattering phase function of asymmetric factor $g$, inclination $i$, and position angle $\theta$. For each disk we created a grid of 500 to 1000 models with $i$, $\theta$, $r_0$ and $g$ as free parameters. 

The models that best fit the observed disks are selected by minimizing the $\chi^2$ value between the NICMOS images and the forward models using the noise maps described in Sec.~\ref{Sec:datared}. Their parameters are reported in Table~\ref{tab:disk}. In the  HD~35650 system, an offset of 20~AU with respect to the star along the disk major axis was additionally needed to model the asymmetry observed in the surface brightness distribution.  This important offset, although sufficiently faithful to the observed disk to calibrate its photometric characteristics, actually reflects a limitation of our modeling process which is adapted to radially symmetric debris disk rather than strongly asymmetric structures as observed around HD~35650. On the other hand, for TWA 25, the difference in brightness between the North and South lobes of the disk is reproduced by our forward model using an axisymmetric disk, suggesting that in this case the apparent asymmetry is a result of the PSF subtraction process. This illustrates that one must be careful when interpreting images after PSF subtraction, for they do not directly reveal the true morphology of disks (cf columns 3 and 4 in Fig.~\ref{Fig:4disks}). This forward modeling process is essential for developing full understanding of the revealed morphologies. Likewise, small-scale structures in the NICMOS images are possible PSF subtraction artifacts and should be interpreted with caution.

The post-processing throughputs are estimated with the ratio of the fluxes of the forwarded models by the fluxes of the unprocessed input models. The total fluxes of disks and their average surface brightnesses  corrected from the post-processing throughput are measured on the unprocessed models and are reported in Table~\ref{tab:disk}.

\begin{deluxetable*}{l|ccccc|ccccc}
\tabletypesize{\footnotesize}
\tablecaption{Disks characteristics\label{tab:disk}}
\tablehead{
\colhead{Name}	&\multicolumn{5}{c}{Best model parameters}    &\colhead{Corrected}	&\colhead{Integrated}			&\colhead{Average}				    	&\colhead{Scattering}        & \colhead{Throughput}	\\
				&\colhead{$r_0$}&\colhead{$i$\tablenotemark{a}}&\colhead{$\theta$\tablenotemark{b}}&\colhead{$g$}&\colhead{$\chi^2_{red}$}	&\colhead{tot. flux}    &\colhead{area}			&	\colhead{SB}					&\colhead{ efficiency}			&\colhead{}\\
				&\colhead{(AU)}&\colhead{($\degr$)}&\colhead{($\degr$)}&\colhead{}&\colhead{}&\colhead{($\mu$Jy)}	        &\colhead{(\arcsec$^2$)}	&\colhead{($\mu$Jy/\arcsec$^2$)}		&\colhead{($\times10^{-4}$)}	&\colhead{(\%)}	
}
\startdata
TWA~7 	(ring: $\alpha_{in}=5$)\tablenotemark{c}	&$35\pm3$		&$22\pm22$		&$59\pm150$	&$0.2_{-0.2}^{+0.8}$     	&1.35	& $523\pm86$		&$4.4\pm1.6$	&$118\pm17$	&$3.7\pm0.6$		&44	\\
TWA~7 	(disk: $\alpha_{in}=0$)\tablenotemark{c}	&$45\pm5$		&$28\pm22$		&$53\pm117$	&$0.1_{-0.1}^{+0.3}$     	&1.28	& $1496\pm130$		&$5.4\pm1.3$	&$276\pm34$	&$10.5\pm0.9$	&27	\\
TWA~25										&$78\pm17$		&$75\pm6$		&$-24\pm8$	&$0.6\pm0.3$		    	&1.24	& $120\pm6$			&$4.3\pm1.1$	&$28\pm5$	&$1.19\pm0.06$	&40	\\
HD~35650 									&$54_{-14}^{+5}$	&$89_{-4}^{+1}$	&$-171\pm7$	&$0.7_{-0.5}^{+0.1}$   	&0.96	& $338\pm15$		&$4.1\pm1.1$	&$82\pm15$	&$0.92\pm0.04$	&55	\\
HD~377 										&$86\pm15$		&$85\pm5$		&$-133\pm4$	&$0.4\pm0.4$		   	&0.59	& $132\pm17$	    	&$3.2\pm1.1$	&$42\pm7$	&$0.30\pm0.04$	&47	\\
\tablenotetext{a}{Inclination is given from face-on.}
\tablenotetext{b}{Orientation of the major axis is given North to East.}
\tablenotetext{c}{\referee{Our modeling of TWA~7 disk shows that the inward power law of its radial profile is not well constrained by our data (difference of 5\% between the $\chi^2_{red}$ values of the best ring model and best disk model). The analysis of the system SED  however indicates that its IR excess is better fitted with two dust populations at different and separated radii each with a single-temperature black-body law  \citep{Riviere-Marichalar2013}. The observed disk is thus more likely a ring (as presented in Fig.~\ref{Fig:4disks}) than a continuous disk.}}
\enddata
\end{deluxetable*}

\section{Discussion}\label{secdisc}

Although protoplanetary disks are as common around low-mass stars as around AFG stars \citep{Andrews2005}, MK stars with debris disks are much  less frequently detected  after the planet forming period as their higher mass cousins.  Two theories can justify these statistics: either low-mass star disks are disrupted shortly after their protoplanetary stage by processes that prevent their transition to evolved debris disks, or MK star disks are more common than we are able to observe but have properties (mass, temperature, composition) that makes them harder to detect. 

On the one hand, while M stars have low radiative pressure that prevents them from blowing small dust grains out of their environments, they are also much more active than higher mass stars, with strong magnetic fields, flares, and stellar winds. M dwarf wind drag force dominates other dynamical mechanisms and can be a major contributor to dust removal from the stellar environment \citep{Plavchan2005}. When formed in high-density clusters, they can also be exposed to strong FUV radiations that could limit planetesimal formation by photoevaporation \citep{Adams2004}. In such dense regions, they are also more subject to close stellar encounters that would strip them from weekly-bound debris disks \citep{Lestrade2011}. 

On the other hand, low-mass stars are intrinsically less luminous than solar-type stars, and their debris disks, if present, would appear colder and fainter in thermal emission. 
\referee{\citet{Andrews2013} showed that protoplanetary disks tend to be less massive around lower-mass stars, dependence that might hold true for their older debris disk counterparts. Other studies also found a trend for less-massive protoplanetary disks to be also smaller \citep{Andrews2010,Pietu2014}, which might also be the case for debris disks too.} 
Combined with unfavorable inclinations, such cold, small, and low-mass disks could indeed lie below the detection limit of current infrared instruments, and be even more critical for resolved imaging where the dust thermal emission is spread on many resolution elements. 
\referee{Although these temperature and mass considerations should not affect detection of low-mass star debris disks in scattered light, their possible smaller size would make their detection more challenging at optical wavelengths too. Finally, dust compositions and grain size distributions are also factors impacting the scattering fraction of debris disks, which might make them harder to detect if different from solar-type disks}. 

Our disk  detections around the low-mass stars TWA~7, TWA~25, and HD~35650 cast a new light on this problem. Unlike AU Mic debris disk, the only other M dwarf disk imaged in scattered light previously, these disks all have a low scattering efficiency (see Table~\ref{tab:disk}, defined by the ratio of the disk flux to the stellar flux) and are relatively close to their host star, a combination which usually makes debris disks difficult to image in scattered light. Yet, our images demonstrate that such low-mass star disks exist and are detectable with modern post-processing techniques. 
TWA~25 is particularly interesting: with only an infrared excess marginally detected in the WISE 3 channel (significance of 4.5), the disk actually seems easier  to detect in scattered light than from its infrared excess, which would indicates a disk with a very low mass and grains with high albedo.

\section{Conclusion}

We presented the detection of four disks imaged for the first time in scattered light around the stars TWA~7, TWA~25, HD~35650, and HD~377. The images were obtained by reanalyzing archival HST-NICMOS data in the F160W and F110W filters, using recently developed PSF subtraction methods. Three of these disks are detected around young M and late-K stars, and form with the disk hosted by AU Mic the only debris disks imaged in scattered light around low mass stars. Unlike AU Mic disk, these three disks appear both faint and compact, which might be a property shared by low-mass star disks that would explain the scarcity of their detections so far. These detection will enable future studies to compare dust properties between MK dwarfs and solar-type stars.

\acknowledgments

This project was made possible by the Mikulski Archive for Space Telescopes (MAST) at STScI  which is operated by AURA under NASA contract NAS5-26555. Support was provided by NASA through grants HST-AR-12652.01 (PI: R. Soummer) and HST-GO-11136.09-A (PI: D. Golimowski), and by STScI Director’s Discretionary Research funds. 

{\it Facilities:} \facility{HST (NICMOS)}.


\end{document}